\documentclass[12pt]{article}
\usepackage[a4paper,margin=1in]{geometry}

\usepackage{floatrow}
\usepackage{amsmath,amssymb}
\renewcommand{\leq}{\leqslant}
\renewcommand{\geq}{\geqslant}

\usepackage{siunitx}
\usepackage{mathtools}  
\usepackage{blkarray} 
\usepackage{array} 

\usepackage{booktabs} 

\usepackage{braket}
\usepackage{graphicx}
\usepackage{microtype}
\usepackage{csquotes} 

\usepackage[svgnames]{xcolor}
\usepackage{color,soul}  

\usepackage{fontspec}

\usepackage{caption}
\usepackage{hyperref} 
\hypersetup{
  pdfauthor={Yanlong Sun},
  bookmarksnumbered,  
  colorlinks=true,      
  linkcolor=blue,       
  citecolor=blue,       
  filecolor=magenta,    
  urlcolor=blue         
}

\usepackage{subcaption}

\usepackage[backend=biber,
  style=phys,  
  biblabel=brackets, 
  maxbibnames=11,  
  doi=true,
  backref=true,
  natbib=true  
]{biblatex}

\DeclareFieldFormat{doi}{%
  \mkbibacro{DOI}\addcolon\space
  \ifhyperref
    {\href{https://doi.org/#1}{\nolinkurl{#1}}}
    {\nolinkurl{#1}}}
\renewbibmacro*{pageref}{%
  \addperiod
  \iflistundef{pageref}
    {}
    {\footnotesize\printtext[parens]{
       \ifnumgreater{\value{pageref}}{1}
         {\bibstring{backrefpages}\ppspace}
     {\bibstring{backrefpage}\ppspace}%
       \printlist[pageref][-\value{listtotal}]{pageref}\addperiod}}}

\newcommand{\ucsection}[1]{%
  \section*{\centering #1}%
  \addcontentsline{toc}{section}{#1}%
}

\let\orgautoref\autoref

\renewcommand{\autoref}{%
  \def\equationautorefname{Eq.}%
  \def\figureautorefname{Fig.}%
  \def\subfigureautorefname{Fig.}%
  \def\sectionautorefname{\S}
  \def\subsectionautorefname{\S}%
  \def\subsubsectionautorefname{\S}%
  \def\Itemautorefname{item}%
  \def\tableautorefname{Table}%
  \def\listingautorefname{Listing}%
  \orgautoref}

\usepackage{xparse}

\DeclareDocumentCommand{\vect}{m o}{\IfNoValueTF{#2}{\mathbf{#1}}{\mathbf{#1}_{#2}}}
\DeclareDocumentCommand{\rset}{m o}{\IfNoValueTF{#2}{\mathbb{#1}}{\mathbb{#1}^{#2}}}
\DeclareDocumentCommand{\cset}{m o}{\IfNoValueTF{#2}{\mathbb{#1}}{\mathbb{#1}^{#2}}}

\newcommand{\norm}[1]{\left\lVert#1\right\rVert} 
\newcommand{\abs}[1]{\left\lvert#1\right\rvert}

\DeclareMathOperator\tr{\mathbf{tr}}
\DeclareDocumentCommand{\spin}{m}{\boldsymbol\sigma \cdot \mathbf{#1}}
\DeclareDocumentCommand{\bket}{m}{\ket{\beta_{#1}}}
\DeclareDocumentCommand{\bbra}{m}{\bra{\beta_{#1}}}
\DeclareDocumentCommand{\corr}{m m m}{#1(\mathbf{#2}, \mathbf{#3})}

\title{\vspace{-1em} The quantum pigeonhole effect as a new form of Bell's theorem without inequality}
\author{
  Yanlong Sun
    \thanks{Institute of Artificial Intelligence, Hefei Comprehensive National Science Center, Hefei, China} 
    \thanks{National Key Laboratory of Human-Machine Hybrid Augmented Intelligence, Xi'an Jiaotong University, Xi'an, China}
  \and
  David H.~Wei 
    \thanks{Quantica Computing LLC, San Jose, CA, USA}
  \and
  Jack W.~Smith
    \thanks{Center for Biomedical Informatics, College of Medicine, Texas A\&M University, Houston, TX, USA}
  \and
  Hongbin Wang \footnotemark[4]
}

\date{\small \today}

\begin{document}

\maketitle

\ucsection{Abstract}

The quantum pigeonhole effect (QPE) appears to contradict the classical pigeonhole principle by allowing three quantum particles distributed between two boxes to exhibit no pairwise coincidence.
We show that this effect does not signal a breakdown of classical counting, but instead arises from quantum contextuality.
By deriving Bell-type inequalities directly from the pigeonhole principle and reformulating the weak-measurement protocol within a bipartite density-operator framework, we demonstrate that the QPE is a form of Bell's theorem without inequalities.
The apparent paradox reflects the impossibility of non-contextual eigenvalue assignments rather than a violation of classical combinatorial logic.

\paragraph{keywords}
Bell's theorem; quantum pigeonhole effect; weak measurement; quantum contextuality; separability

\section{Introduction}

The pigeonhole principle, also known as Dirichlet's drawer principle \cite{Dirichlet1863}, is a basic result in combinatorics and number theory \cite{Allenby2011}.
It asserts that placing three objects into two boxes necessarily forces at least two objects to occupy the same box.
The quantum pigeonhole effect (QPE), however, appears to challenge this classical reasoning:
for suitably prepared and post-selected quantum systems, three particles distributed between two boxes can exhibit instances in which no two particles are observed in the same box \cite{Aharonov2016}.
The QPE arises from weak measurements performed on pre- and post-selected ensembles \cite{Aharonov1988,Marcovitch2007,Aharonov2013-PRA,AharonovY2013Cheshire}.
Although experimentally demonstrated \cite{WaegellM2017pra,ChenMC2019pnas,LiuZH2020natcomm}, its interpretation remains controversial \cite{AharonovY2021pla,KunstatterG2020pla,KunstatterG2021pla,CorreaR2021pra}.
The effect therefore raises questions not only about the ontological status of quantum states, but also about the logical structure underlying classical counting principles.

Since the formulation of the Einstein–Podolsky–Rosen (EPR) paradox \cite{EPR1935} and Bell's theorem \cite{Bell1964}, such paradoxes have been increasingly understood through the framework of quantum contextuality.
Quantum contextuality is the impossibility of extending eigenvalue assignments consistently across overlapping commuting sets of observables.
While functional relations among operators hold within each measurement context, no global valuation preserves these additive or multiplicative constraints across all contexts simultaneously.
Originally formalized by Kochen and Specker \cite{KochenS1967} and independently analyzed by Bell \cite{BellJS1966rmp}, contextuality has since been developed in operational and structural forms \cite{PeresA1991jpa,SpekkensW2005pra,CabelloA2008prl,CabelloA2014prl,AbramskyS2011sheaf,AbramskyS2024rsta-intro}, and has found applications beyond quantum mechanics in psychology, cognition, and large language models \cite{DzhafarovEN2016book,AbramskyS2024rsta,LoKI2025rspa}.

In this work, we show that the QPE does not invalidate the classical pigeonhole principle.
Rather, the apparent contradiction stems from implicitly assuming non-contextual value assignments.
We derive two Bell-type inequalities directly from the pigeonhole principle and demonstrate that the weak-measurement scheme of \cite{Aharonov2016} can be reformulated within a bipartite density-operator framework.
Our results indicate that the QPE is best understood as a form of Bell's theorem without inequalities.
The QPE paradox is thus not a failure of counting, but a failure of global non-contextual description.

\section{Bell's Inequalities by the Pigeonhole Principle}
We first show how the pigeonhole principle, when coupled with non-contextual reasoning, can give rise to two new forms of Bell's inequality.

\begin{figure}[!ht]
\centering
\begin{subfigure}{0.26\textwidth}
  \raisebox{15pt}{\includegraphics[width = \textwidth]{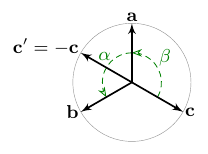}}
  \caption{}
  \label{fig:qpe-Bell-a}
\end{subfigure}\hspace{20pt}
\begin{subfigure}{0.36\textwidth}
  \includegraphics[width = \textwidth]{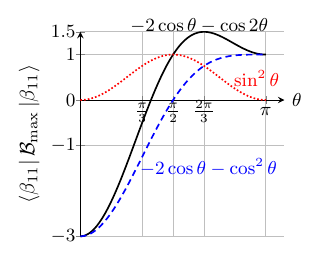}
  \caption{}
  \label{fig:qpe-Bell-b}
\end{subfigure}
\caption{Bell's inequalities by the pigeonhole principle.
(\textbf{A}): Mermin's version of the EPR paradox, and the equivalence between the Bell's inequalities {\protect \autoref{eq:Bell-ineq-upper} and \autoref{eq:Bell-original-v2}}.
(\textbf{B}): When measuring the singlet $\bket{11}$, the total correlation $\bbra{11} \mathcal{B}_{\max} \bket{11}$ (solid black), and its components along $Z \otimes Z$ (dashed blue) and $X \otimes X$ (dotted red), as the functions of detector orientations ({\protect \autoref{eq:Bell-Op-max-Bell}}).
The maximal violation of {\protect \autoref{eq:Bell-ineq-upper}} is observed when $\theta_{\vect{a}, \vect{b}} = \theta_{\vect{b}, \vect{c}} = \theta_{\vect{c}, \vect{a}} = 120 \si{\degree}$, and the maximal violation of {\protect \autoref{eq:Bell-original-v2}} is observed when $\theta_{\vect{a}, \vect{c}'} = \theta_{\vect{b}, \vect{c}'} = 60 \si{\degree}$, and $\theta_{\vect{a}, \vect{b}} = 120 \si{\degree}$.
}
\label{fig:qpe-Bell}
\end{figure}

We start with David Mermin's representation of the EPR paradox and Bell's theorem \cite{Mermin1981ajp,Mermin1985} \cite[367]{PenroseR1999emperor} (\autoref{fig:qpe-Bell-a}).
A pair of spatially separated particles are sent to two distant observers.
Each observer can independently choose one of three measurement settings, corresponding to spin measurements along three different directions $\vect{a}$, $\vect{b}$, and $\vect{c}$, and the measurement outcomes are represented by three binary random variables $a, b, c = \pm 1$, respectively.
Let $\corr{\rho}{x}{y} = \braket{xy}$ denote the correlation between $\vect{x}$ and $\vect{y}$, where $\braket{xy}$ is the expected value of the products $xy$.
By the pigeonhole principle, at least two of the variables take the same sign, so that at least one of the products $ab$, $ac$, and $bc$ is equal to $+1$ (perfect correlation), and $ab + ac + bc \geq -1$.
If we assume that the measurement outcome is independent from how the measurement is performed, then
\begin{equation}\label{eq:Bell-ineq-lower}
  \corr{\rho}{a}{b} + \corr{\rho}{a}{c} + \corr{\rho}{b}{c} \geq -1.
\end{equation}

Similarly, if the measurement outcomes are respectively $a,b,c = \pm i$ (here $\pm i$ represent two orthogonal states that can be measured with a Hermitian operator with corresponding eigenstates), then the pigeonhole principle dictates that at least one of the products $ab, bc, ac$ is equal to $-1$ (perfect anti-correlation), so that $ab + ac + bc \leq 1$, and
\begin{equation}\label{eq:Bell-ineq-upper}
  \corr{\rho}{a}{b} + \corr{\rho}{a}{c} + \corr{\rho}{b}{c} \leq 1.
\end{equation}

Apparently, \autoref{eq:Bell-ineq-lower} and \autoref{eq:Bell-ineq-upper} belong to the family of Bell's inequalities.
To see the equivalence, the original Bell's inequality \cite{Bell1964} 
was given in the form
\begin{equation}\label{eq:Bell-original}
  1 + \corr{\rho}{b}{c'} \geq \abs{\corr{\rho}{a}{b} - \corr{\rho}{a}{c'}}.
\end{equation}
Rearrange,
\begin{equation}\label{eq:Bell-original-v2}
  \corr{\rho}{a}{b} - \corr{\rho}{a}{c'} - \corr{\rho}{b}{c'} \leq 1,
\end{equation}
which is equivalent to \autoref{eq:Bell-ineq-upper} if $\vect{c'}=-\vect{c}$ or $c'=-c$ (\autoref{fig:qpe-Bell-a}).

More generally, let $a, b, a', b' = \pm 1$ or $\pm i$, so that $a(b + b') + a'(b - b') = \pm 2$, leading to the CHSH inequality \cite{CHSH1969},
\begin{equation}\label{eq:CHSH-corr}
  -2 \leq \corr{\rho}{a}{b} + \corr{\rho}{a}{b'} + \corr{\rho}{a'}{b} - \corr{\rho}{a'}{b'} \leq 2.
\end{equation}
Let $a' = -b$ and $b'=c$, \autoref{eq:CHSH-corr} reduces to \autoref{eq:Bell-ineq-lower} if $b=\pm 1$ and reduces to \autoref{eq:Bell-ineq-upper} if $b= \pm i$.

However, quantum contextuality forbids the observable properties of a system from being predefined independently from how they are observed.
Suppose $\vect{a}$, $\vect{b}$, and $\vect{c}$ are in the same plane at equal intervals $\theta = 120 \si{\degree}$.
If the two particles have been prepared in the Bell state $\bket{00}= \tfrac{1}{\sqrt{2}}\big(\ket{00} + \ket{11}\big)$, then quantum mechanics predicts that the correlation between the settings is $\corr{\rho}{x}{y} = \cos 2 \theta$, so that
\begin{equation}\label{eq:Bell-ineq-lower-violated}
  \corr{\rho}{a}{b} + \corr{\rho}{a}{c} + \corr{\rho}{b}{c} = 3 \cos 2 \theta = -\tfrac{3}{2},
\end{equation}
which violates \autoref{eq:Bell-ineq-lower} by a factor of $1.5$.
If the two particles have been prepared in the Bell state $\bket{11}= \tfrac{1}{\sqrt{2}}\big(\ket{01} - \ket{10}\big)$, then we have $\corr{\rho}{x}{y} = -\cos \theta$, and
\begin{equation}\label{eq:Bell-ineq-upper-violated}
  \corr{\rho}{a}{b} + \corr{\rho}{a}{c} + \corr{\rho}{b}{c} = -3 \cos \theta = \tfrac{3}{2},
\end{equation}
which violates \autoref{eq:Bell-ineq-upper} by a factor of $1.5$ as well (\autoref{fig:qpe-Bell-b}).
We will show that both are maximal violations by the eigen-analysis of Bell operators in \autoref{eq:Bell-Op-max-Bell}, then show the maximal violation of the CHSH inequality in \autoref{eq:Bell-Op-max-CHSH}.

Note that Mermin's account of the EPR paradox does not explicitly invoke the pigeonhole principle, nor does it question classical counting methods. 
Instead, it emphasizes the logical impossibility of assigning context-independent local values (so-called ``instruction sets'') to quantum observables.
Subsequent unified proofs make clear that quantum contextuality emerges from the impossibility of extending additive or multiplicative eigenvalue assignments consistently across overlapping commuting contexts \cite{Bell2004,PeresA1990pla,Mermin1990PRL,Mermin1993}.

It is also worth noting that Mermin's treatment in \cite{Mermin1981ajp} influenced Feynman \cite[366]{Feynman2005Perfectly}, who later reformulated the three-setting configuration as a six-setting scheme in his seminal 1982 paper \cite{Feynman1982}, linking the EPR paradox to the conceptual foundations of quantum computation.
Feynman's argument highlights nonclassical correlations through the probability of agreement, namely, the likelihood that measurements performed under two different settings produce identical outcomes.
In our case here, when measuring photon spins prepared in $\bket{00}$, the probability of agreement between two settings is $P^=_{\vect{x,y}} = \cos^2 \theta$, so that the quantum correlation is $\corr{\rho}{x}{y} = 2 P^=_{\vect{x,y}}  - 1 = \cos 2 \theta$ (\autoref{eq:Bell-ineq-lower-violated}).

\section{The Bipartite Quantum Pigeonhole Effect}
At the heart of the quantum pigeonhole effect (QPE) is the disappearance of entanglement of a bipartite system, where a part of the post-selected state in a tripartite system is orthogonal to a Bell state at the intermediate time.
We will call this effect the bipartite quantum pigeonhole effect (BQPE).

Consider two particles to be put in two boxes labeled by $0$ and $1$.
We first prepare each particle in an even superposition $\ket{+} = \frac{1}{\sqrt{2}} (\ket{0} + \ket{1})$ so that the pre-selected state $\ket{\psi} = \ket{+}\ket{+}$ is a product state that contains no correlation.
Construct two projectors,
\begin{equation}\label{eq:Aha-big-Pi}
  \Pi^{\mathrm{same}} = \ket{00}\bra{00} + \ket{11}\bra{11}, \quad \Pi^{\mathrm{diff}} = \ket{01}\bra{01} + \ket{10}\bra{10},
\end{equation}
where $\Pi^{\mathrm{same}}$ finds two particles in the same box, $\Pi^{\mathrm{diff}}$ finds them in different boxes.
Given that $\Pi^{\mathrm{same}} + \Pi^{\mathrm{diff}} = I$ and $\braket{\psi | \Pi^{\mathrm{same}} | \psi}  = \braket{\psi | \Pi^{\mathrm{diff}} | \psi} = \tfrac{1}{2}$, the states immediately after the projective measurements are two maximally entangled Bell states,
\begin{align}\label{eq:BQP-Bell}
\begin{split}
  \frac{\Pi^{\mathrm{same}} \ket{\psi}}{\sqrt{\braket{\psi | \Pi^{\mathrm{same}} | \psi}}}
  &=\tfrac{1}{\sqrt{2}}\big(\ket{00} + \ket{11}\big) = \bket{00}, \\
  \frac{\Pi^{\mathrm{diff}} \ket{\psi}}{\sqrt{\braket{\psi | \Pi^{\mathrm{diff}} | \psi}}}
  &=\tfrac{1}{\sqrt{2}}\big(\ket{01} + \ket{10}\big) = \bket{01}.
\end{split}
\end{align}
Let $\ket{\phi_k}$ denote the post-selected states as the direct products of the orthogonal states $\ket{+i} = \frac{1}{\sqrt{2}} \big(\ket{0} + i\ket{1}\big)$ and $\ket{-i} = \frac{1}{\sqrt{2}} \big(\ket{0} - i\ket{1}\big)$,
\begin{align}\label{eq:Aha-post-bipartite}
\begin{split}
  \ket{\phi_1} &= \ket{+i} \ket{+i} , \quad  \ket{\phi_2} = \ket{+i} \ket{-i} , \\
  \ket{\phi_3} &= \ket{-i} \ket{+i} , \quad  \ket{\phi_4} = \ket{-i} \ket{-i} ,
\end{split}
\end{align}
where $\sum_k \ket{\phi_k} \bra{\phi_k} = I$.
Then, the probabilities of detecting $\bket{00}$ (two particles in the same box) and $\bket{01}$ (two particles in different boxes) are respectively,
\begin{align}\label{eq:Aha-vanish-eff-bipartite}
\begin{split}
  \abs{\braket{\phi_1 |\beta_{00}}}^2 &= \abs{\braket{\phi_4 |\beta_{00}}}^2
  = \abs{\braket{\phi_2 |\beta_{01}}}^2 = \abs{\braket{\phi_3 |\beta_{01}}}^2 = 0, \\
  \abs{\braket{\phi_2 |\beta_{00}}}^2 &= \abs{\braket{\phi_3 |\beta_{00}}}^2
  = \abs{\braket{\phi_1 |\beta_{01}}}^2 = \abs{\braket{\phi_4 |\beta_{01}}}^2 = \tfrac{1}{2}.
\end{split}
\end{align}

This is all possible combinations of BQPE with orthogonal states $\ket{+i}$ and $\ket{-i}$.
To see how it extends to QPE, construct the pre-selected state for a tripartite system as $\ket{\Psi} = \ket{+}_1 \ket{+}_2 \ket{+}_3$, and the post-selected state as $\ket{\Phi} = \ket{+i}_1 \ket{+i}_2 \ket{+i}_3$.
Apply the intermediate measurement $\Pi^{\mathrm{same}}$ in \autoref{eq:Aha-big-Pi} on particles $1$ and $2$, then 
\begin{align}\label{eq:Aha-Eq7-corrected}
\begin{split}
  \bra{\Phi} \big(\Pi^{\mathrm{same}}_{1,2} \otimes I \big) \ket{\Psi}
  &=\big(\bra{+i}_1 \bra{+i}_2 \Pi^{\mathrm{same}}_{1,2} \ket{+}_1 \ket{+}_2 \big) \braket{+i | +}_3 \\
  &=\tfrac{1}{\sqrt{2}} \braket{\phi_1 | \beta_{00}} \braket{+i | +}_3 =0.
\end{split}
\end{align}
The entanglement in $\bket{00}$ vanishes in the post-selection $\braket{\phi_1 | \beta_{00}}$, which means that we cannot find particles $1$ and $2$ in the same box.
Let $\lambda_i$ be the eigenvalues for the final locations of particle $i$, we have $\lambda_1 \neq \lambda_2$.
We may also obtain $\lambda_2 \neq \lambda_3$ and $\lambda_1 \neq \lambda_3$ by applying $\Pi^{\mathrm{same}}$ to every other two particles.
However, in each measurement, only two particles are measured, and to the best of our knowledge, the third particle remains in the superposition $\ket{+} = \frac{1}{\sqrt{2}} (\ket{0} + \ket{1})$.
QPE becomes a paradox only when we assume that a particle has a definite value $\lambda_i$ without being measured.
It contradicts to the classical pigeonhole principle only when we attempt to compare all $\lambda_i$ at the same time.

\section{Bell's Theorem without Inequality}
\autoref{eq:Aha-Eq7-corrected} indicates that QPE is in fact a new version of Bell's theorem without inequality, in the same vein as the Greenberger-Horne-Zeilinger (GHZ) state \cite{GHSZ1990,MerminND1990GHZ} and Hardy's state \cite{Hardy1993,Mermin1995BestBell}.
Although the original Bell's theorem lacks the elements of weak measurement, it differs from BQPE and QPE only in the ``directions'' of entanglement detection.
For instance, tracing out one of the qubits in a GHZ state may leave the remaining state completely unentangled \cite{Dur2000}.
With a non-detected third particle, a Bell state violates maximally the Bell inequalities but their mixture with equal weights exhibits no violation \cite{Popescu1992,HorodeckiR1996-two-spin-half}.
In the framework of entanglement witness \cite{HorodeckiM1996,Leinaas2006,Branciard2013}, apparent paradoxes are resolved by discriminating entangled states from separable states.

The directions of entanglement detection can cleanly defined by the Pauli representation of density operators.
Reformulate the post-selected states in \autoref{eq:Aha-vanish-eff-bipartite} with a pair of density operators  $\varrho^{\pm}_Y$,
\begin{align}\label{eq:Aha-post-bipartite-do}
\begin{split}
  \varrho^+_Y &= \tfrac{1}{2} \big(\ket{\phi_1} \bra{\phi_1} + \ket{\phi_4} \bra{\phi_4} \big)
        = \tfrac{1}{2} \big(\bket{01} \bbra{01} + \ket{\beta_{10}} \bra{\beta_{10}}\big), \\
  \varrho^-_Y &= \tfrac{1}{2} \big(\ket{\phi_2} \bra{\phi_2} + \ket{\phi_3} \bra{\phi_3} \big)
        = \tfrac{1}{2} \big(\bket{00} \bbra{00} + \ket{\beta_{11}} \bra{\beta_{11}}\big),
\end{split}
\end{align}
where $\bket{00}= \tfrac{1}{\sqrt{2}}\big(\ket{00} + \ket{11}\big)$, $\bket{01}= \tfrac{1}{\sqrt{2}}\big(\ket{01} + \ket{10}\big)$, $\bket{10}= \tfrac{1}{\sqrt{2}}\big(\ket{00} - \ket{11}\big)$, and $\bket{11}= \tfrac{1}{\sqrt{2}}\big(\ket{01} - \ket{10}\big)$.
The operators $\varrho^{\pm}_Y$ have a curious property:
Each of them is an equal mixture of two maximally entangled Bell states, yet at the same time a \emph{separable state} that contains no quantum correlation. 
By the orthonormality of Bell states,
\begin{align}\label{eq:vanishing-trace}
\begin{split}
  \tr \big(\varrho^{+}_Y \bket{00} \bbra{00}\big) &= \tr \big(\varrho^{-}_Y \bket{01} \bbra{01}\big) = 0, \\
  \tr \big(\varrho^{-}_Y \bket{00} \bbra{00}\big) &= \tr \big(\varrho^{+}_Y \bket{01} \bbra{01}\big) = \tfrac{1}{2} \, .
\end{split}
\end{align}
This is exactly the same result as \autoref{eq:Aha-vanish-eff-bipartite}, where we obtain a vanishing trace when a post-selected state $\ket{\phi_k}$ is orthogonal to a Bell state.
We can find the entire family of $\varrho^{\pm}_Y$ in all directions in terms of Pauli matrices,
\begin{align}\label{eq:post-family}
\begin{split}
  \varrho^+_Z &= \tfrac{1}{2} \big(\bket{00} \bbra{00} + \bket{10} \bbra{10} \big) = \tfrac{1}{4} (I \otimes I + Z \otimes Z), \\
  \varrho^+_X 
    &= \tfrac{1}{2} \big(\bket{00} \bbra{00} + \ket{\beta_{01}} \bra{\beta_{01}}\big) = \tfrac{1}{4} (I \otimes I + X \otimes X), \\
  \varrho^+_Y 
    &= \tfrac{1}{2} \big(\bket{01} \bbra{01} + \ket{\beta_{10}} \bra{\beta_{10}}\big) = \tfrac{1}{4} (I \otimes I + Y \otimes Y), \\
  \varrho^-_Y 
    &= \tfrac{1}{2} \big(\bket{00} \bbra{00} + \ket{\beta_{11}} \bra{\beta_{11}}\big) = \tfrac{1}{4} (I \otimes I - Y \otimes Y), \\
  \varrho^-_X 
    &= \tfrac{1}{2} \big(\bket{10} \bbra{10} + \ket{\beta_{11}} \bra{\beta_{11}}\big) = \tfrac{1}{4} (I \otimes I - X \otimes X), \\
  \varrho^-_Z 
    &= \tfrac{1}{2} \big(\bket{01} \bbra{01} + \ket{\beta_{11}} \bra{\beta_{11}}\big) = \tfrac{1}{4} (I \otimes I - Z \otimes Z).
\end{split}
\end{align}
The transformations between these states are either unitary (by Hadamard gate $H \otimes H$ or phase gate $S \otimes S$), or partial transposition ($\varrho^-_Y  = (T \otimes I) \varrho^+_Y$), so that separability is preserved  \cite{Peres1996,HorodeckiM1996}.
If we write Bell states as,
\begin{equation*}
  \bket{ab} \bbra{ab}
  = \tfrac{1}{4}\big(I \otimes I + (-1)^b Z \otimes Z + (-1)^a X \otimes X - (-1)^{a+b} Y \otimes Y \big),
\end{equation*}
then all scenarios of vanishing trace can be summarized as
\begin{equation}\label{eq:tr-vanishing}
  \tfrac{1}{4} \tr\big((\sigma_i \otimes \sigma_i)(\sigma_j \otimes \sigma_j)\big) 
  = \delta_{ij} \, ,
\end{equation}
where $\sigma_0 = I$, $\sigma_1 = X$, $\sigma_2=Y$, and $\sigma_3=Z$.
For example, the $Y \otimes Y$ component in $\varrho^+_Y$ coupling with the $-Y \otimes Y$ component in $\bket{00} \bbra{00}$ annihilates the trace in $I \otimes I$, whereas cross terms such as $(Y \otimes Y)(X \otimes X)$ remain traceless.

For the Bell inequalities in \autoref{eq:Bell-ineq-lower} and \autoref{eq:Bell-ineq-upper}, following \cite{HorodeckiR1995-violating-Bell}, we can define a Bell operator,
\begin{equation}\label{eq:Bell-op}
  \mathcal{B} = \spin{a} \otimes \spin{b} + \spin{a} \otimes \spin{c} + \spin{b} \otimes \spin{c} \, ,
\end{equation}
where $\spin{a} = \sum a_i \sigma_i $, $\sum_i a_i^2 = 1$, $\corr{\rho}{a}{b} = \braket{\spin{a} \otimes \spin{b}}$, and so on.
Since cross terms remain traceless, we set $\vect{a} = (0,0,1)$, $\vect{b} = (\sin \alpha , 0 , \cos \alpha)$, and $\vect{c} = (-\sin \beta, 0 , \cos \beta)$ in the same plane, where $\alpha, \beta \in [0, \pi]$ (\autoref{fig:qpe-Bell-a}).
Then, \autoref{eq:Bell-op} is reduced to
\begin{align}\label{eq:QPE-Bell-op-trimmed}
\begin{split}
  \mathcal{B} &= (\cos \alpha + \cos \beta + \cos \alpha  \cos \beta )(Z \otimes Z) - \sin \alpha \sin \beta \, (X \otimes X).
\end{split}
\end{align}
The minimum of $\bbra{00} \mathcal{B} \bket{00}$ and the maximum of $\bbra{11} \mathcal{B} \bket{11}$ are found at $\alpha = \beta = 120 \si{\degree}$, or, $\theta_{\vect{a}, \vect{b}} = \theta_{\vect{b}, \vect{c}} = \theta_{\vect{c}, \vect{a}} = 120 \si{\degree}$ (\autoref{fig:qpe-Bell-b}).
This gives us the Bell operator that maximally violates the Bell inequalities in \autoref{eq:Bell-ineq-lower} and \autoref{eq:Bell-ineq-upper},
\begin{equation}\label{eq:Bell-Op-max-Bell}
  \mathcal{B}_{\max} = -\tfrac{3}{4} (Z \otimes Z + X \otimes X),
\end{equation}
with eigenpairs $\big(-\tfrac{3}{2}, \bket{00}\big)$, $\big(0, \bket{01}\big)$, $\big(0, \bket{10}\big)$ and $\big(\tfrac{3}{2}, \bket{11}\big)$.
The same bounds $\pm \tfrac{3}{2}$ have been found by \cite{Ardehali1998PRA}, but our method here is directly from the eigenvalues of Bell operators.
Substituting $-\vect{c'}$ for $\vect{c}$ in \autoref{eq:Bell-op} gives us the maximal violation of  the original Bell inequality in \autoref{eq:Bell-original-v2} at $\theta_{\vect{a}, \vect{c}'} = 60 \si{\degree}$, $\theta_{\vect{b}, \vect{c}'} = 60 \si{\degree}$, and $\theta_{\vect{a}, \vect{b}} = 120 \si{\degree}$ (\autoref{fig:qpe-Bell-a}).

\begin{figure}[!ht]\centering
\includegraphics[scale=0.88]{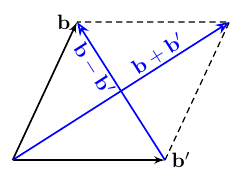}
\caption[QPE (rhombus)]{The maximal violation of the CHSH inequality ({\protect \autoref{eq:CHSH-corr}}) is observed when $\vect{b} \cdot \vect{b'} =0$, and $\max (\norm{\vect{b} + \vect{b'}} + \norm{\vect{b} - \vect{b'}}) = 2\sqrt{2}$ ({\protect \autoref{eq:Bell-Op-max-CHSH}}).}
\label{fig:QPE-rhombus}
\end{figure}

To get a better picture on how violations of Bell inequalities arise from quantum contextuality, we consider the more general CHSH inequality in \autoref{eq:CHSH-corr}, particularly regarding the sum of eigenvalues.
The Bell operator for the CHSH inequality is given by \cite{HorodeckiR1995-violating-Bell},
\begin{equation}\label{eq:CHSH-contrast}
  \mathcal{B}_{\mathrm{CHSH}} = \spin{a} \otimes \spin{(b+b')} + \spin{a}' \otimes \spin{(b-b')}.
\end{equation}
Each of the observables $\spin{b}$ and $\spin{b}'$ has eigenvalues $\pm 1$, but $\spin{(b+b')}$ has eigenvalues $\pm \norm{\vect{b} + \vect{b'}}$, and $\spin{(b-b')}$ has eigenvalues $\pm \norm{\vect{b} - \vect{b'}}$, where $\norm{\star}$ is the Euclidean norm.
Crucially, by Weyl's inequalities, the eigenvalues of a sum of Hermitian matrices are not, in general, equal to the sums of the eigenvalues of the individual matrices \cite{KnutsonTao2001}.
As illustrated in \autoref{fig:QPE-rhombus}, $\norm{\vect{b} + \vect{b'}}$ and $\norm{\vect{b} - \vect{b'}}$ are not independent, and the maximum of $\norm{\vect{b} + \vect{b'}} + \norm{\vect{b} - \vect{b'}}$ is reached when $\vect{b} \cdot \vect{b'} =0$.
This gives us the Bell operator that maximally violates the CHSH inequality,
\begin{equation}\label{eq:Bell-Op-max-CHSH}
  \mathcal{B}^{\mathrm{CHSH}}_{\max} = \sqrt{2} (Z \otimes Z + X \otimes X),
\end{equation}
with eigenpairs $\big(2\sqrt{2}, \bket{00}\big)$, $\big(0, \bket{01}\big)$, $\big(0, \bket{10}\big)$ and $\big(-2\sqrt{2}, \bket{11}\big)$.
The largest and smallest eigenvalues $\pm 2 \sqrt{2}$ are known as the Tsirelson bound \cite{Tsirelson1980}.

A comparison of \autoref{eq:post-family}, \autoref{eq:Bell-Op-max-Bell} and \autoref{eq:Bell-Op-max-CHSH} makes clear that the vanishing trace in QPE and the violations of Bell's inequalities are two sides of the same coin, as they only differ in the directions of the eigenstates of corresponding observables.

\section{Conclusion}

The quantum pigeonhole effect (QPE) has been widely interpreted as a challenge to classical combinatorial reasoning.
In this work, we have shown that the apparent violation of the pigeonhole principle does not reflect a breakdown of classical counting logic, but rather the impossibility of extending non-contextual eigenvalue assignments across incompatible measurement contexts.
By deriving Bell-type inequalities directly from the pigeonhole principle and reformulating the weak-measurement protocol within a bipartite density-operator framework, we have demonstrated that the QPE is structurally equivalent to Bell's theorem without inequalities.
The paradox thus arises from contextual constraints intrinsic to quantum theory, not from arithmetic inconsistency.

This perspective resonates with recent discussions emphasizing the interpretational structure of quantum mechanics.
In particular, as highlighted in the account of quantum Bayesianism (QBism) \cite{Fuchs2014QBism,Mermin2014QBism,SavitskyZ2025sci}, quantum theory may be understood as a framework governing an agent's probabilistic expectations about measurement outcomes rather than as a catalogue of pre-existing properties.
From such a viewpoint, the QPE does not signal that particles ``avoid sharing boxes'', but that joint value assignments across incompatible contexts are not meaningful.
The effect reflects limits on global description rather than violations of classical logic.

Rather than introducing a new departure from established principles, the quantum pigeonhole effect therefore reinforces a familiar lesson: quantum theory constrains how measurement outcomes can be coherently related across contexts.
In this sense, the QPE stands alongside Bell's theorem and Kochen–Specker contextuality as another manifestation of the structural limitations of non-contextual reasoning.

\clearpage

\printbibliography[heading=bibintoc,title={References}]

\end{document}